\documentstyle[preprint,aps]{revtex}

\newcommand{\referencestyle}{
\small
\abovedisplayskip=6pt
\belowdisplayskip=6pt
\vspace{12pt}}

\def\D{\Delta}

\def\t0{\tau_0}

\def\ben{\begin{eqnarray}}
\def\enn{\end{eqnarray}}
\def\ov{\over\displaystyle\strut}
\def\dst{\displaystyle\strut}
\def\l({\left(}
\def\r){\right)}
\def\mn{\langle n \rangle}
\def\mnc{\langle n_c \rangle}
\def\mnh{\langle n_h \rangle}

\begin{document}
\rightline{LUNFD6/(NFFL-7088)-Rev. 1994}
\rightline{hep-ph/9411307}
\begin{center}
{ \Large\bf
Bose-Einstein Correlations for \\
Systems with Large Halo}
\end{center}
\medskip
\begin{center}
        T. Cs\"org\H o$^{1,2}$\footnote{E-mails:
                csorgo@sunserv.kfki.hu, $\qquad$ bengt@quark.lu.se,
                $\qquad$ zimanyi@rmki.kfki.hu }
        B. L\"orstad$^{2,*}$ and J. Zim\'anyi$^{1,*}$
\end{center}
\medskip
\begin{center}
{\it
$^1$ MTA KFKI RMKI, H--1525 Budapest 114, P.O. Box 49. Hungary \\
$^{2}$Department of Elementary Particle Physics, Department of Physics, \\
University of Lund,
Professorsgatan 1,
S - 223 63 Lund, Sweden
}
\end{center}
\vfill
\begin{abstract}
We discuss
Bose-Einstein correlation functions and momentum distributions
which describe boson-emitting systems containing a central part
surrounded by a large  halo.
If the characteristic size of the halo is sufficiently
large, the contributions of central part and the halo to the
 invariant momentum distribution are shown to be
separable.  The momentum-dependence of the intercept
parameter of the correlation function plays
a central role.
 Almost all high energy reactions including lepton-lepton,
lepton hadron, hadron-hadron and nuclear reactions are
interpretable as boson emitting systems with large halo.
The results are applied to
high energy heavy ion data taken
 at CERN SPS. New insights are obtained for the description of
 low transverse momentum enhancement of pions.
\end{abstract}
\vfill
\rightline{Z. Phys. C in press}

 \vfill\eject

\underline{\it Introduction.} Hanbury-Brown -- Twiss
 correlations were discovered 40 years
ago, revealing information about the angular diameters of distant
stars~\cite{HBT}. The method, also referred to as intensity interferometry,
extracts information from the quantum statistical correlation function
for (partially) chaotic fields. The method has extensively been applied
to the study of the freeze-out geometry in high energy nucleus-nucleus
collisions as well as in elementary particle reactions. For recent reviews
see refs.~\cite{bengt,zajc}.

Accumulating evidence indicates that the space-time structure of the pion
emission
in heavy ion reactions at the 200 AGeV bombarding energy region at CERN SPS
has a peculiar feature, namely that the boson emission can be approximately
 divided into two parts: the centre and the
halo~\cite{halo1,halo2,halo3,halo4}.
The central part corresponds to a direct production mechanism
e.g. hydrodynamic evolution or particle production from excited strings,
followed by subsequent re-scattering of the particles.
It is surrounded by a pionic halo which corresponds to pions emitted from the
decay of
long-lived hadronic resonances, like $\omega, \eta$, $\eta'$ and $K^0$,
which have a mean decay length of more than 20 fm.
These resonances give rise to mainly low transverse momentum pions.
The kaonic halo is very rare, created by the weak decays of charmed mesons.
The  halo for $\gamma$ production is more populated,  since some
photons are emitted
from the decays of abundant long lived resonances like $\pi^0$, $\eta$
and  $\eta'$.

Within the halo of long lived resonances,
the momentum distribution of the emitted bosons is independent
from the (approximate)
position of the decay. This in turn is smeared out over a large
region due to the long mean decay length.
This mechanism will result in bosons of similar momentum from
a very large volume. This is in contrast to the core which when
expanding fast in a hydrodynamic manner will have pions of similar momentum
from a relatively small volume only.

The key point is the following:
Let us consider an ensemble of long-lived resonances
with similar momentum, emitted from a given small volume of the core.
The momentum distribution of the {\it decay products} of these resonances
will be similar to each other, independently of the approximate position
of the decay. Now the approximate position of the decay is randomly
distributed along the line of the resonance propagation with the
weight $P(t) \propto \exp(- m_{res} \Gamma_{res} t /E_{res} )$.
Thus the decay products will be emitted with the same momentum
distribution from a volume which is
elongated along the line of resonance propagation, given by
$V_{decay} \simeq A_0 \mid p_{res} \mid /(m_{res} \Gamma_{res})$,
where $A_0$ is the initial transverse size of the surface
through which the resonances are emitted with a momentum $p_{res}$
approximately at the time of the decay of the core, $\tau_s$.

Note that  the volume of this elongated tube which emits resonance
decay products with certain momentum is not only proportional
to $1/\Gamma_{res}$ but it depends also on the momentum of the
resonance in form of a factor of $\mid {\bf p} \mid /m$. Thus the
considered volume is large only if the  decaying resonances
satisfy $\mid {\bf p} \mid / (m \Gamma_{res})  >> 1$
fm/c. This in turn implies that
the simple picture that the halo can be identified with the
decay products of long-lived resonances may need further corrections
for pions at rapidity $ y = 0$ with low transverse momentum.

Note that there is a gap in the life-time distribution
of the abundant hadronic resonances which decay to pions:
$ 1 / \Gamma_{\rho } \simeq 1.3$ fm/c,
$1/\Gamma_{N*} \simeq 0.56$ fm/c,
$ 1/\Gamma_{\Delta } \simeq 1.6 $ fm/c and the relatively rare
$K^*$ has a life-time of $ 1/\Gamma_{K*} \simeq 3.9 $ fm/c,
while each of the long-lived resonances has a life-time
bigger than 20 fm/c: $1/\Gamma_{\omega} = 23.5$ fm/c,
$1/\Gamma_{\eta^{\prime}} = 986.5 $ fm/c, $1/\Gamma_{\eta} = 164 400 $
fm/c, $ 1/\Gamma_{K_s^0} = 2.76 $ cm/c.
 Thus the life-times of the long-lived hadronic resonances
are at least a factor of 5 - 10 longer than the life-times
of the short-lived hadronic resonances. The life-times
of short-lived resonances
are of the same
order of magnitude than the time-scales
for re-scattering at the time of the last hadronic interactions.
Thus the decay-products of the short-lived resonances will mainly
contribute to the core, which is resolvable by BEC measurements,
while the decay-products
of long-lived hadronic resonances will mainly
belong to the halo, re-defined alternatively as
the part of the emission function which is not resolvable
in a given  Bose-Einstein measurement.

In the present paper we investigate the phenomenological consequences
of such a structure where there is a characteristic length-scale
of the core which is smaller than any length-scale that can
 be connected
to the surrounding halo. Halo length-scales larger than 20 fm
 give rise to a sharp peak of the correlation function
in the $Q \le \hbar/R \approx 10$
MeV/c region, where $Q$ stands for the measured component of the relative
momentum.
This region is strongly influenced by the Coulomb interaction thus the
accuracy of the correlation function determination will strongly depend on the
Coulomb corrections which constitute a problem in itself~\cite{zinov}.
Furthermore, this
$Q < 10$ MeV/c region is very difficult to measure due to
the circumstance that we must properly
determine two very close tracks
which makes experimental systematic errors the largest
in this region of the correlation function even for high resolution experiments
like
the NA35 and NA44 high energy heavy ion experiments at CERN.

We will show in this paper that even if we cannot measure reliably the
correlation function below
$Q<10$ MeV/c important new information about the particle production in a
core/halo
scenario can be given by analyzing both the invariant momentum distribution of
single particles
and the correlation function for $Q>10$ MeV/c. We will apply our approach to
NA44 data on
$S + Pb$ reactions at 200 AGeV/c.

In the present paper we investigate the core/halo scenario
 utilizing analytical results in the Wigner-function formalism.  This
formalism is
well suited to describe both the invariant momentum distribution (IMD) and
the Bose-Einstein correlation function (BECF). We utilize the version of the
formalism
discussed in refs.~\cite{pratt_csorgo,zajc} and applied to analytic
calculations
in refs.~\cite{lutp,nr,1d,3d,uli_s,uli_l} recently.

\underline{\it Review of Wigner-function formalism.} In the Wigner-function
formalism,
the one-boson emission is characterized by the emission function, $S(x,p)$,
which
can be considered as the time derivative of the
(non-relativistic) single-particle Wigner function~\cite{lutp}
or as the covariant Wigner-transform of the source density matrix,
{}~\cite{pratt_csorgo,zajc,chapman_heinz}. Here
$x = (t,{\bf r}\,) = (t, r_x, r_y, r_z)$ denotes the four-vector in space-time,
and $p = (E, {\bf p} \,) = (E, p_x, p_y, p_z)$ stands for the four-momentum
of the on-shell particles with mass $m = \sqrt{ E^2 - \bf p^{\, 2} }$.
The transverse mass is defined as $m_t = \sqrt{m^2 + p_x^2 + p_y^2}$
and $y = 0.5 \log\l({\dst E+p_z \ov E-p_z} \r) $ stands for the rapidity.

For chaotic sources the two-particle Wigner-function can be expressed
in terms of the symmetrized products of the single-particle Wigner functions.
Based on this property, the two-particle Bose-Einstein correlation functions
are determined from the single-particle  emission function solely.
These relations are especially simple in terms of the Fourier-transformed
emission function,
\ben
\tilde S(\Delta k , K ) & = & \int d^4 x \,\,
                 S(x,K) \, \exp(i \Delta k \cdot x ),\label{e:ax}
\enn
where
\begin{eqnarray}
\Delta k  = p_1 - p_2, & \quad \mbox{\rm \phantom{and}}\quad &
K  = {\displaystyle\strut p_1 + p_2 \ov 2}
\end{eqnarray}
and $\Delta k \cdot x $ stands for the inner product
of the four-vectors.

The {\it inclusive} $l$-particle invariant momentum distribution
of the emitted particles reads as
\ben
N_l({\bf p}_1,...,{\bf p}_l) & = & {1\ov \sigma_{in}} E_1 ...  E_l
{\dst d \sigma \ov d {\bf p}_1 \,  ... \, d {\bf p}_l },\label{e:incl}
\enn
where $\sigma_{in}$ is the total inelastic cross section.
These inclusive $l$-particle momentum distributions are normalized to
the $l$-th factorial moments of the multiplicity distribution as
\ben
\langle n (n-1) ... (n -l + 1) \rangle & = &
\int ... \int  {d {\bf p}_1 \ov E_1}  ... {d {\bf p}_l \ov E_l}
\, N_l(p_1, ..., p_l)
\enn
These distributions are basically the IMD-s of particle numbers (IMD-N).

The absolute normalization of the
IMD-s (or the corresponding emission
functions) is to be considered carefully.
The IMD-s of the detection probability (IMD-P-s)
\ben
        P_1({\bf p}_1) & = & { 1 \ov \langle n \rangle} \,
                        {E_1 \ov  \sigma_{in}} \,
                        {\dst d\sigma \ov d {\bf p}_1 }, \\
        P_2({\bf p}_1,\, {\bf p}_2)
                 & = & { 1 \ov \langle n(n-1) \rangle} \,
                        {E_1 E_2 \ov  \sigma_{in}} \,
        {\dst d^{6} \sigma \ov d {\bf p}_1 d {\bf p}_2 },
\enn
are also frequently used
in the literature, (e.g. see refs.~\cite{pratt_csorgo,nr,1d}),
which are normalized to unity as
\ben
\int {\dst d {\bf p}_1 \ov E_1} \, P_1({\bf p}_1)
                \,\, & = & 1,\label{e:nimd} \\
\int {\dst d {\bf p}_1 \ov E_1} {\dst d {\bf p}_2 \ov E_2} \,
                P_2({\bf p}_1,{\bf p}_2)
                \,\, & = & 1 . \label{e:nimd2}
\enn

For cylindrically symmetric systems, the probability distribution
in $y$ and $m_t$ variables, $d^2 n / dy / d(m_t^2)$ shall also be utilized.
These distributions are integrals of the $P_1(\bf p)$ distribution over the
polar angle $\phi$ in the transverse plane. Obviously, one has
$d^2 n / dy /d(m_t^2) = \pi P_1({\bf p}(y,m_t,\phi_0))$, where $\phi_0$
is an arbitrary polar angle and $P_1$ is independent of $\phi_0$.

In the versions of the Wigner-function formalism presented in
refs.~\cite{pratt_csorgo,nr,1d}, the phase-space integral
of the emission function was normalized to 1.
This is to be contrasted to the versions of the hydrodynamic
descriptions of the IMD-s and BECF-s in refs.~\cite{uli_s,uli_l,aver,QM95}
where the phase-space integral of the emission function is normalized to
$\mn $. For the sake of clarity, we distinguish these
emission functions by the subscripts $_w$ (for Wigner) and $_{hyd}$
for hydrodynamics. Obviously, their relation is given by
\ben
        S_{hyd}(x,p) & = & \mn S_w(x,p).
\enn
The one-particle invariant momentum distributions (IMD-P-s and
IMD-N-s) may be expressed with the emission function as
\ben
        P_1(\bf p) & = & \tilde S_w(\Delta k = 0, K = p),\label{e:imdp} \\
        N_1(\bf p) & = & \tilde S_{hyd}(\Delta k = 0, K = p).\label{e:imdn}
\enn
The two-particle
 BECF-s are expressed in terms of our  auxiliary
function, $\tilde S_w(\Delta k, K)$ as
\ben
C(K,\Delta k) &  = & {\dst \langle n \rangle^2 \ov \langle n (n-1) \rangle}
\, {\dst N_2( {\bf p}_1 , {\bf p}_2)   \ov
        N_1( {\bf p_1} ) \, N_1({\bf p}_2 ) }
                = {\dst P_2({\bf p}_1, {\bf p}_2 ) \ov P_1({\bf p}_1)
                        P_1({\bf p}_2) } \\
\null &  = &    1 +  {\displaystyle\strut
         \mid \tilde S_w(\Delta k , K) \mid^2 \ov
                \tilde S_w(0,p_1) \, \tilde S_w(0,p_2) }
 \simeq   1 +  {\displaystyle\strut
         \mid \tilde S_w(\Delta k , K) \mid^2 \ov \mid \tilde S_w(0,K)\mid^2 },
\enn
 as was presented e.g. in ref. \cite{pratt_csorgo,zajc,nr}.
The corrections to the last approximation are known to be negligible
{}~\cite{uli_l}.  Effects of final state  Coulomb and Yukawa interactions
are neglected as implicitly assumed by the above equations,
and completely chaotic  emission is assumed.
The formulas recollected up to this point shall be utilized in the
subsequent part of this paper.

{\it \underline{A core and halo model.} }
The Wigner-functions are in general complex valued functions, being the
quantum analogue of the classical phase-space distribution functions.
  For the purpose of Monte-Carlo simulations of Bose-Einstein correlation
functions the off-shell Wigner-functions are approximated by the on-shell
classical emission functions obtained in a simulation~\cite{pratt_csorgo,RQMD};
 in case of analytic
calculations they are usually modeled by some positive valued
functions~\cite{lutp,nr,1d,3d,uli_s,uli_l}.
In this paper, neither of these approximations seems to be necessary.
In fact we need only the following simplifying assumptions:

{\it Assumption 1}. The bosons are emitted either from a
central {\it core} or
from the surrounding  {\it halo}. The respective emission functions
are indicated by $S_c(x,p)$ and $S_h(x,p)$, and $f_c $ indicates the
 fraction of the bosons emitted from the central part.
According to this assumption, the complete emission function can be written
 as
\ben
 S_w(x,p) = f_c  S_{c}(x,p) + ( 1 - f_c) S_{h}(x,p).
\enn
Both the emission function of the centre and that of the halo are normalized
to unity here, accordingly to the Wigner-function normalization.

{\it Assumption 2}. We assume that the emission function $S_h(x,p)$, which
characterizes the halo, changes on a scale $R_h$ which is larger
 than  $R_{max}\approx \hbar / Q_{min}$, the maximum length-scale resolvable
by the intensity interferometry microscope.
In our considerations,
 $Q_{min}$ is taken as 10 MeV/c.
However, the smaller central part
of size $R_c$ is
assumed to be resolvable,
\ben
 R_h > R_{max} > R_c.
\enn
This inequality is assumed to be satisfied by all characteristic scales
in the halo and in the central part, e.g. in case the side, out or longitudinal
components~\cite{bertsch,lutp} of the correlation function are not identical.

As mentioned in the introduction, the halo
of long-lived resonances is characterized by large regions of homogeneity
{}~\cite{aver,uli_l} according to this Assumption.
This is motivated by the fact that
particles emerging from decays of long-lived resonances are emitted
with a momentum distribution independent of the approximate position of the
decay, which in turn is smeared out over a large volume being proportional
to the  decay-time of the (long-lived) resonances.

Note, that {\it Assumption 2} relies on

{\it Assumption 0}: The core does {\it not} have a self-similar, no-scale
structure.
  The length-scale(s) of the core is (are) resolvable by the
HBT microscope.

Based on {\it Assumption  1} the IMD-s can be re-written as
\ben
 P_1({\bf p}) = f_c \, P_{1,c}({\bf p}) + (1 - f_c) \, P_{1,h} ({\bf p}) ,
        \label{e:ich}
\enn
where the subscripts $c,h$ index the contributions by the central and the
halo parts  to the IMD given  respectively by
\ben
  P_{1,i}({\bf p}) & = & \tilde S_i(\Delta k = 0, K = p)
\enn
for $i = c,h$.
Note that the IMD as expressed by eq.~(\ref{e:ich}) includes the possibility
that the IMD for the bosons of the halo is different from the bosons emerging
from the central part. Thus the relative contribution of the halo and the
centre is a function of the momentum in this model.

The normalization condition according to {\it Assumption 1} reads as
\ben
\int {\dst d{\bf p} \ov E} \, P_{1,i}({\bf p}) \,\, & = & 1\label{e:nimdi}
\enn
for $i = c,h$.
At this point the normalization conditions described for the hydrodynamic
and Wigner emission functions can be utilized.
According to Assumption 1, the hydrodynamic emission-function is a sum, too:
\ben
        S_{hyd}(x,p) & = & \mn \, f_c \, S_c(x,p) +
                                \mn \, (1-f_c) \, S_h(x,p).\label{e:sh}
\enn
It is convenient to define the 'hydrodynamic' emission function for the core
and the halo,
\ben
        S_{hyd,c}(x,p) & = & \mn \, f_c \, S_c(x,p), \label{e:shc}\\
        S_{hyd,h}(x,p) & = & \mn \, ( 1- f_c) \, S_h(x,p). \label{e:shh}
\enn
Since these functions are normalized to the mean multiplicity  of the core
and the halo, respectively, it follows that
\ben
        \mnc & = & f_c \mn , \label{e:mnc} \\
        \mnh & = & (1 - f_c) \mn. \label{e:mn}
\enn
Thus the core fraction $f_c$ turns out to be the the ratio of the mean
multiplicity
of the core to the mean multiplicity of the whole system.

Integrating eqs.~(\ref{e:sh}-\ref{e:shh}) over the space-time variables
and using eq.~(\ref{e:imdn}),
one obtains that
\ben
        N_1({\bf p}) & = & N_{1,c}({\bf p}) + N_{1,h}({\bf p}),
\enn
and as a consequence of eq.~(\ref{e:incl}) one finds that
\ben
        E {\dst d\sigma \ov d{\bf p} } & = &
                E {\dst d\sigma_c \ov d{\bf p} } +
                E {\dst d\sigma_h \ov d{\bf p} }.
\enn
In the forthcoming it shall be shown that the above cross-sections,
describing particle production cross-sections from the core and the halo,
can be related to measurable quantities.

The 'ideal' shape for the BECF is expressed by
\ben
C(K,\Delta k) = 1 +
 {\dst \mid f_c \, \tilde S_c(\D k, K) + ( 1-f_c) \, \tilde S_h(\D k, K) \mid^2
\ov
       \mid f_c \, \tilde S_c(\D k=0, K=p) + ( 1-f_c) \, \tilde S_h(\D k=0,
K=p) \mid^2 },
        \label{e:cmat}
\enn
which includes interference terms for boson pairs of $(c,c)$ $(c,h)$ and
$(h,h)$ type.
Due to the assumption that the emission is completely chaotic, the exact value
of the BECF at  $\mid {\bf \D k} \mid = 0$ is always 2 in this model.

The measured two-particle BECF is determined for
$\mid {\bf \D k} \mid > Q_{min}\approx 10$ MeV/c,
and any structure within the $ \mid {\bf \D k} \mid < Q_{min}$ region is not
resolved.
However, the $(c,h)$ and $(h,h)$ type boson pairs create a narrow peak
in the BECF exactly in this $\D k $ region according to eq.~(\ref{e:cmat}),
which cannot
be resolved according to {\it Assumption 2}. From this assumption,
$\tilde S_h(\Delta k, K) \simeq 0 $ for $\mid \Delta {\bf k} \mid > Q_{min}$ or
for $\mid \Delta k^0 \mid  > Q_{min}$. With other words,
 the Fourier-transformed emission function of the halo for non-zero
relative momenta vanishes at the given resolution $Q_{min}$.
At zero relative momentum the same quantity gives the single particle
momentum
distribution from the halo, which is not effected by the {\it two}-particle
resolution.

Including the finite resolution effect, symbolized by the horizontal bar
over the correlation function,
 the measured BECF can be written as
\ben
\overline{ C(\D k, K) } & = & 1 +
      \lambda_*(K=p;Q_{min}) \, R_c(\D k, K),
\enn
where the effective intercept parameter and the correlator of the core are
defined as
\ben
\lambda_*(K=p;Q_{min})& = & {\dst \mid f_c \tilde S_c( \D k=0, K=p) \mid^2 \ov
                      \mid f_c \, \tilde S_c(\D k=0, K=p) +
                 ( 1-f_c) \, \tilde S_h(\D k=0,K=p)\mid^2},
                           \label{e:lamq} \\
R_c(\D k, K) & = &              {\dst \mid \tilde S_c( \D k, K) \mid^2 \ov
                             \mid \tilde S_c( \D k = 0, K=p) \mid^2}.
\enn
This  expression indicates that the only effect of a large resonance halo
is to introduce a momentum-dependent effective intercept parameter
to the two-particle correlation function, while the relative momentum
dependent factor shall coincide with the correlator $R_c(\D k, K)$ of the core.
The effective intercept parameter can in turn be expressed with the IMD-P-s as
\ben
\lambda_*(K=p;Q_{min}) =  f_c^2 {\dst P_{1,c}^2({\bf p}) \ov P_1^2({\bf p}) }.
        \label{e:14}
\enn
Alternatively, the effective intercept
$\lambda_*(K=p;Q_{min})$ can be rewritten with the help of
 eq. (\ref{e:mnc}) as
\ben
\lambda_*(K=p;Q_{min}) =
        \left[ {\dst N_{1,c}({\bf p}) \ov N_{1}({\bf p}) } \right]^2 =
         \left[  {\dst d\sigma_c \ov d{\bf p}}
                        \left( {\dst d\sigma \ov d{\bf p}} \right)^{-1}
\right]^2.
\enn

In the forthcoming, we shall utilize the notation
$ \lambda_*({\bf p} ) = \lambda_*(K=p;Q_{min})$.
The effective intercept parameter $\lambda_*({\bf p} )$
shall thus in general  depend on the
mean momentum of the observed boson pair,
and we have $0 \le \lambda_*({\bf p})
 \le 1$.
The equation above shows that this
effective $\lambda_*({\bf p} )$ has a very simple interpretation
as the square of the fraction of core particles to all particles emitted
with momentum ${\bf p}$.
We also see that this equation makes it possible to measure the cross-section
of the bosons produced
in the central core
or from the halo
 with the help of the combined use of the BECF parameter $\lambda_*({\bf p})$
and the cross-section of all particles.
For the sake of completeness, we list these relations below:
\ben
        {\dst d\sigma_c \ov d{\bf p}} & = &
                  \, \sqrt{\lambda_*({\bf p}) } \, \, {\dst d\sigma \ov d{\bf
p}}, \\
        {\dst d\sigma_h \ov d{\bf p}} & = &
                \, ( 1 - \sqrt{\lambda_*({\bf p}) } ) \, \,
                {\dst d\sigma \ov d{\bf p}}.
\enn
Alternatively, if a calculation predicts $N_c({\bf p} )$ from a model of
the core, this prediction can be combined with the measured $ \lambda_*({\bf p}
)$
to describe the IMD-N-s for the whole system as
\ben
        N({\bf p} ) & = & {\dst 1 \ov \sqrt{ \lambda_*({\bf p} )} }
N_{1,c}({\bf p}) .
\enn
Note that the $\lambda_*({\bf p})$ effective intercept parameter in our
picture reflects the drop in the BECF
from its exact value of $\lambda_{xct} = 2$  due to the contribution from the
resonance halo, when measured with the finite two-particle momentum
resolution
$Q_{min}$. Thus the measured intercept,
$\lambda_*({\bf p})$ does {\it not} coincide with the
exact value of the BECF at zero relative momentum
within this description, see Figure 1 for illustration.

The effect of the halo is to introduce
 a momentum-dependent effective intercept parameter to the
BECF. The shape of the BECF for
$Q>Q_{min}$ shall be solely determined by the freeze-out
phase-space distribution in the central part.
 This central part or core is usually well accessible
to hydrodynamic calculations, see
 refs.~\cite{aver,nr,1d,3d,uli_s,uli_l} for example.

Note also that the presented results
 have in principle
nothing to do with the similar relation obtained for the
case of partially coherent, partially chaotic fields, where $\lambda_* =
f_{inc}^2$,
and $f_{inc} = \langle n_{inc} \rangle /
\langle n_{tot} \rangle $ gives the fraction of the total
multiplicity
in the chaotic field, although the relation formally is similar.
For the case of partially coherent fields, the Bose-Einstein
correlation function contains an interference term in between the coherent and
the chaotic
fields which leads to a double-Gaussian or double-exponential
structure~\cite{gyu_ka},
emphasized recently in ref.~\cite{weiner}.
In our case, the measurable part of the BECF contains not three but only two
terms,
and it looks very much like the BECF would look
like for the case the halo were missing i.e. $f_c = 1$.

Note also, that the focus is {\it not} on the reduction of the intercept of the
BECF
due to the halo (resonance decays) but more precisely on the {\it
momentum-dependence}
of this reduction. The fact of the reduction has been known before~
\cite{halo2,halo1,pratt_csorgo,RQMD}, however its momentum dependence was not
utilized as far as we know. Within our formalism, it is straightforward
to show that the core fraction of bosons can be estimated by
\ben
f_c & = & \int {\dst d{\bf p} \ov E} \, \sqrt{\lambda_*(\bf p)} \,\,
         P_1({\bf p}) .
 \label{e:d}
\enn
Alternatively, $f_c$ can be expressed as
\ben
        f_c & =  &{\dst \mnc \ov \mn},
\enn
where
\ben
        \mnc & = &\int {\dst d{\bf p} \ov E}
        \, \sqrt{\lambda_*(\bf p)} \,\,
                N_1({\bf p}).
\enn
{}From eq.(\ref{e:d}) it follows that the inequalities
\ben
 \min \sqrt{\lambda_*({\bf p})}\, \le \, f_c \, \le  \,
  \max \sqrt{\lambda_*({\bf p})}
\enn
are also satisfied. Please note that the above formulas are also valid
in the special case of the constant function,
$\lambda_*({\bf p}) = const$.

\underline{\it Application}.
Present NA44 data~\cite{glasgow} for central $S + Pb$ reactions at CERN SPS
with
200 AGeV bombarding energy show an approximately $m_t$ independent  intercept
parameter:
$\lambda_{\pi^+} = 0.56 \pm 0.02$ and $0.55 \pm 0.02$ at the quite different
mean
transverse momenta of $150 $ MeV and $ 450 $ MeV
\footnote{
Note that  with increasing $ \langle p_t \rangle$
the NA44 acceptance moves slightly  in rapidity, too.
For  pions in NA44 acceptance
$ \langle p_t \rangle = 150 $ MeV corresponds to $\langle y \rangle = 3.5$
while $\langle  p_t \rangle = 450 $ MeV corresponds to $\langle y \rangle =
2.7$ both values being within 0.5 units from mid-rapidity.
That's why we expect the change in $m_t$ be the dominant effect.
}, respectively. This suggests
that
in the mid-rapidity region, where the NA44 data were taken,
 the momentum distribution of pions from the resonance halo is
similar to that from
the central part, and the halo contains $ 1 - \sqrt{\lambda_*}
=
25 \pm 2 $ \% of all the pions.
{}From eq.~(\ref{e:14}) we can see that this observed constancy
of $\lambda_*$ in the NA44 acceptance region\footnote{
The observed momentum-independence of the parameter $\lambda(y.m_t)$
is further supported by  the NA35 collaboration
(which has a wide acceptance both in $y$ and $m_t$).
NA35 collaboration
also finds  an approximately $y$ and $m_t$ independent $\lambda(y,m_t)$
parameter too, as can be seen on Figs. 3 a) and b) in T. Alber et al,
Z. Phys. {\bf C66} (1995) 77.
}
leads to the following simple equalities
for the IMD and the BECF:
\ben
{\dst d^2 n \ov dy \, dm_t^2 } \, & = & \, {\dst d^2 n_c \ov dy \, dm_t^2 }
                                \, = \, {\dst d^2 n_h \ov dy \, dm_t^2 },\\
\overline{ C(\D k, K) } \,  & = & \, 1 + f_c^2 R_c(\D k, K).
\enn
I.e. the only apparent effect of the halo is to reduce the intercept
parameter of the measured BECF to $\lambda_*  = f_c^2$ while the IMD and
the $(\D k, K)$ dependence of the BECF is determined by the central
part {\it exclusively}.

The resonance halo was expected to play a more active role:
Pions coming from the resonance-halo surrounding a strongly
interacting centre are predominant at lower values of the transverse momentum
according to the SPACER calculations in ref.~\cite{lutp} and RQMD calculations
in ref.~\cite{RQMD}, similarly to
the results of calculations with HYLANDER~\cite{halo2} and the hydrodynamic
calculations of the Regensburg group~\cite{uli_h}. The resonance fraction
as a function of the transverse momentum was explicitly shown in the
publications
{}~\cite{RQMD,halo2}. Each of the mentioned models predicts that the halo of
resonances produces predominantly low transverse
momentum pions, and they predict that the effective
 intercept $\lambda_*(p;Q_{min})$ should increase with increasing
transverse mass in contrast to the constancy of the  measured values.

The constancy of the effective parameter $\lambda_*(y,m_{t})$ with respect to
$m_{t}$ suggests a mechanism of enhancing the low momentum particles in the
core.
Such an enhancement has a natural explanation in a hydrodynamic description of
an expanding core as we have presented in \cite{QM95}, further
elaborated in a forthcoming paper~\cite{3d}. The mechanism is a simple volume
effect, see also \cite{uli_l}.
 The calculated cross-section is proportional to
 the volume where the particles
with a given momentum are emitted from. This
(effective and momentum-dependent) volume is
 measured by the BECF. In the NA44 experiment
all three radii of this volume are {\it measured}
  and found to be equal  within the errors of the measurement.
  Further, all the three radius components are found to be
inversely proportional the the square root of $m_{t}$~\cite{glasgow}.
This will give an effective volume factor,
$V_* \propto (m_{t})^{-3/2}$, enhancing the single particle
cross-section of the core at low $m_{t}$.
This is to be contrasted to the case of a static fire-ball,
 which has a cross-section
proportional to  $V$, the (momentum-independent) volume of the fire-ball.
This effect is illustrated on Figure 2. Note that the enhancement
at low transverse momentum
(as compared to the static source)
is a factor of 10 larger for pions than kaons or omega mesons,
due to the relatively small rest mass of pions. In case of pions,
the  concave shape of the momentum distribution
is clearly seen on Figure 2, for heavier particles the concave
shape is less detectable. To our best knowledge, this
transverse mass dependent effective volume factor
$V_* \propto (m_{t})^{-3/2}$, which results in a factor of 10
enhancement of low-$p_t$ pions from the core, as compared to low-$p_t$
heavy resonances from the core, has not yet been taken into account in
any previous numerical analysis of low-transverse momentum enhancement.

\underline{\it Discussion.} The general result for the BECF of systems with
large halo
 coincides with the most frequently applied phenomenological parameterizations
 of the BECF in high energy heavy ion as well as
in high energy particle reactions~\cite{bengt}.
Previously, this form has received a lot of criticism from the theoretical
side,
claiming that it is in disagreement with quantum statistics~\cite{weiner}
or that the $\lambda $ parameter is just a kind of fudge parameter,
 a measure of our ignorance,
which has been introduced to make theoretical predictions comparable to data.
Now one can see that this type of parameterizations can be derived
with a standard inclusion of quantum statistical effects -- if we assume that
we discuss interferometry for systems with large halo. Most of
the studied reactions including $e^+ + e^-$ annihilations
various lepton-hadron and hadron-hadron reactions nucleon-nucleus and
nucleus - nucleus collisions are phenomenologically well
describable~\cite{bengt} with variations of eq.~(12).
This fact does not exclude the possibility
that each boson emitting system created in the above mentioned
 high energy reactions contains a large halo.

Let us remark that there is a standard experimental procedure  to check the
sensitivity of the fitted Bose-Einstein correlation function parameters to the
variation of the bin-size. The data point in the first bin is frequently
left out from the fitting since this data point is
unreliable~\cite{bengt,glasgow}.
These add up to a check of the stability of the fitted
$\lambda_*(y,m_t) $ parameter(s)
for the variation of $Q_{min}$, the size of the excluded region.
Clear separation of the emission
function into a core and a halo part is possible in a given reaction only if
this or a similar experimental test indicates the insensitivity of the
fitted parameters to the exact value of $Q_{min}$, which may vary
in a certain small relative momentum region around 10 MeV/c.

The applicability of the halo picture to
a given reaction is not necessarily the only possible explanation
of a reduced intercept. It is known that the final state interactions
may have an influence on the effective intercept of the two-particle
correlation functions~\cite{bowler} and the effect has been shown to
result in a stronger drop of the intercept value for smaller source
i.e. for hadronic strings created in lepton-lepton, lepton-hadron
or hadron-hadron collisions. However, more detailed calculations
indicate that the higher order corrections for the final state interactions
may very well cancel the first order effects resulting in a very small
total final state interaction correction~\cite{bow}.
In our calculations these effects have been neglected, so our results
should in principle be compared to data which present the {\it genuine}
Bose-Einstein correlation function. This function is not necessarily the
same as the short-range part of the two-particle correlation function
since Coulomb and Yukawa interactions as well as the $\eta' \eta$ decay chain
and other short-range correlations may mask the quantum statistical
correlation functions. However, as the particle density is increased
(in the limit of very large energy or large colliding nuclei),
the Bose-Einstein correlations together with the final state interactions
 dominate over other short-range correlations
due to combinatoric reasons.

It is well known that the hydrodynamic predictions ~\cite{yuri,aver}
for the rapidity and the transverse mass dependence of the longitudinal radius
parameters of the Bose-Einstein correlation functions are found to
describe the available data in heavy ion reactions at CERN SPS to the best
available precision~\cite{NA35,glasgow}, although the calculations were
performed
within a hydrodynamic framework, and the effects caused by resonance decays
were neglected in the calculation. The intercept parameter
was  $\lambda_{th} = 1$ in  the calculation, while the intercept parameter
in the NA35 and NA44 experiments was significantly lower than that.
Thus the scenario discussed in this paper supports such a possibility,
since it has been shown that  a large halo of long-lived resonances
results in {\it i)} a momentum-dependent effective intercept parameter
 $\lambda_*({\bf p})$ and {\it ii})
a special relative momentum dependence of the measurable part of the
 Bose-Einstein correlation function,
 which is prescribed by the emission function of the
core.

\underline{\it In summary}, we have studied the case when
 the central boson-emitting region
is surrounded by a large halo,
which also emits bosons. If the size of the halo is
so large that it cannot be resolved in Bose-Einstein correlation measurements,
 lot of information shall be concentrated in the momentum dependence of the
intercept parameter of the correlation function.
We have shown that with the help of the Bose-Einstein correlation
measurement, the invariant momentum distribution can be measured for the
two independent components
belonging to the core and the halo, respectively.
The results do not depend on any particular parametrization of the core
nor of the halo.

Analysis of the NA44 data for two-pion correlations
indicated that the normalized invariant momentum distribution
of the pions from the halo of long-lived resonances
within errors coincides with the normalized invariant momentum
distribution of the pions from the central core.
This result can be explained with a hydrodynamic model of the core
evolution.
The number of pions coming from the halo region, characterized by
large length-scales of $ R_h \ge 20$ fm,
 was found to be $ 25 \pm 2$ \%
of the total number of pions within the NA44 acceptance.

{\it Acknowledgments:}
Cs.T. would like to thank B. L\"orstad
for special hospitality at University of Lund,
G. Gustafson, S. Hegyi, S. S. Padula and M. Pl\"umer
  for stimulating discussions on the subject,
M. Gyulassy and
Xin-Nian Wang for kind hospitality at Columbia University and at
Lawrence Berkeley Laboratory.
This work has been supported
by the Human Capital and Mobility (COST) programme of the
EEC under grant No. CIPA - CT - 92 - 0418
(DG 12 HSMU), by the Hungarian
NSF  under Grants  No. OTKA-F4019, OTKA-T2973 and
OTKA-W/01015107
and by the USA-Hungarian Joint  Fund
by grant  MAKA 378/93.

\vfill
\eject
\begin{center}
{\bf Figure captions}
\end{center}

{\bf Figure 1}. The shape of the BECF is illustrated for a source containing a
core
and a large halo. The contribution from the halo is restricted to the shaded
area,
while the shape of the BECF outside this interval is determined completely
by the contribution of the core. If the resolution for a given experiment
is restricted to $Q > 10 $ MeV, then an effective and momentum dependent
intercept parameter, $\lambda_*(y,m_t)$ will be measured, which can be combined
with the measured momentum distribution to determine the the momentum
distribution
of the particles emitted directly from the core.

{\bf Figure 2}. The  momentum
   distributions $d^2N/ (m_t^2 \, dy \, dm_t)$
   are shown  for pions, kaons and omega mesons in
    arbitrary units as a function of $m_t - m$, when the
    effective $m_t$ dependent volume factor, $V_* \propto
    1/m_t^{(3/2)}$ is taken into account.
    To illustrate the effect of the volume factor $V_*$ an effective
temperature
    $T_{eff} = 200 $ MeV was used for each particle, and we utilized
    $m_{\pi} = 140 $ MeV, $m_K = 494 $ MeV and  $m_{\omega} = 782$ MeV.
    The plotted functions are all proportional to
    ${\dst 1 \ov m_t^{(3/2)}} \exp(- (m_t - m) / T_* )$, being
    normalized to have the same value at $m_t - m = 2.0 $ GeV.
    Note that the enhancement at low transverse momentum
    is a factor of 10 larger for pions (solid line)  emitted
    from the core than the enhancement for the heavier omega mesons (dashed
line).
        The kaons (short-dashed line) and omega mesons on the other hand
        have similar momentum distributions.
        Dotted line stands  for the  momentum distribution of a static
        source, which is proportional to $\exp( - m_t/T_*)$.
\vfill\eject


\begin{thebibliography}{99}
\referencestyle
\bibitem{HBT}            R. Hanbury-Brown and R. Q. Twiss, Phyl. Mag.
        {\bf 45} 663 (1954); R. Hanbury-Brown and R. Q. Twiss, Nature
        (London) {\bf 177}, (1956) 27 and {\bf 178}, (1956) 1046
\bibitem{bengt}          B. L\"orstad, Int. J. Mod. Phys.
                        {\bf A12} (1989) 2861
\bibitem{zajc}          W. A. Zajc in "Particle Production in Highly Excited
                        Matter", ed. by H. Gutbrod and J. Rafelski,
                        {\bf NATO ASI series B303} (Plenum Press, 1993) p. 435
\bibitem{halo1}         T. Cs\"org\H o et al, Phys. Lett. {\bf B241} (1990) 301
\bibitem{halo2}         J. Bolz et al,  Phys. Rev. {\bf D47} (1993) 3860
\bibitem{halo3}         J. Bolz et al,  Phys. Lett. {\bf B300} (1993) 404
\bibitem{halo4}         B. R. Schlei et al,  Phys. Lett.{\bf B293} (1992) 275
\bibitem{zinov}         D. Anchiskin and G. Zinovev, preprint BI-TP-94-19,
                        Phys. Rev. {\bf C51} (1995) R2306
\bibitem{pratt_csorgo}  S. Pratt, T. Cs\"org\H o and J. Zim\'anyi,
                        Phys. Rev. {\bf C42} (1990) 2646
\bibitem{lutp}          T. Cs\"org\H o and S. Pratt, {\bf KFKI-1991-28/A}, p.
75
\bibitem{nr}            T. Cs\"org\H o, B. L\"orstad and J. Zim\'anyi,
                        Phys. Lett. {\bf B338}  (1994) 134;
                        T. Cs\"org\H o, M. Asakawa, J.Helgesson and B.
L\"orstad,
                        nucl-th/9506006, CU-TP-697, LBL-37305,
LUNFD6/(NFFL-7111) 1995
\bibitem{1d}            T. Cs\"org\H o,  Phys. Lett. {\bf B347} (1995) 354
\bibitem{3d}            T. Cs\"org\H o and B. L\"orstad,
                        "Bose-Einstein Correlations for Three-Dimensionally
                        Expanding, Cylindrically Symmetric Finite Systems",
                        preprint LUNFD6/(NFFL-7082)-Rev. 1994, hep-ph/9509213
\bibitem{uli_s}         S. Chapman, P. Scotto and U. Heinz,
                         hep-ph/ 9408207, Phys. Rev. Lett. {\bf 74} (1995) 4400
\bibitem{uli_l}         S. Chapman, P. Scotto and U. Heinz,
                        hep-ph/ 9409349, Heavy Ion Physics {\bf 1} (1995) 1
\bibitem{chapman_heinz} S. Chapman and U. Heinz,
			Phys. Lett. {\bf B340 } (1994) 250
\bibitem{aver}          V. A. Averchenkov, A. N. Makhlin and Yu. M. Sinyukov,
                        Yad. Fiz. {\bf 46} 1525-1534 (1987),
                        Sov. J. Nucl. Phys. {\bf 46} 905 (1987)
\bibitem{QM95}           T. Cs\"org\H o and B. L\"orstad,
                        {\bf A590} (1995) 468c, hep-ph/9503494
\bibitem{RQMD}          J. P. Sullivan et al, Phys. Rev. Lett. {\bf 70} (1993)
3000
\bibitem{bertsch}        G. F. Bertsch, Nucl. Phys. {\bf A498} (1989) 173c
\bibitem{gyu_ka}        M. Gyulassy, S. K. Kaufmann and L. W. Wilson,
                        Phys. Rev. {\bf C20}, (1979) 2267
\bibitem{weiner}        R.M. Weiner, Phys. Lett. {\bf B232} (1989) 278
\bibitem{glasgow}       H. Beker et al., NA44 Collaboration,
                        Phys. Rev. Lett. {\bf 74} (1995) 3340
\bibitem{uli_h}         J. Sollfrank et al, Z. Phys.{\bf C52} (1991) 593;
                        J. Sollfrank et al,  Phys. Lett. {\bf  B252} (1990) 256
\bibitem{bowler}        M. G. Bowler, Z. Phys. {\bf C39}, (1988) 81
\bibitem{bow}           M. G. Bowler, Part. World {\bf 2} (1991) 1-6
\bibitem{yuri}          A. Makhlin and Y. Sinyukov,
                        Z. Phys. {\bf C39} (1988) 69
\bibitem{NA35}          Th. Alber et al, NA35 collaboration,
                        Phys. Rev. Lett. {\bf 74} (1995) 1303
\end{thebibliography}
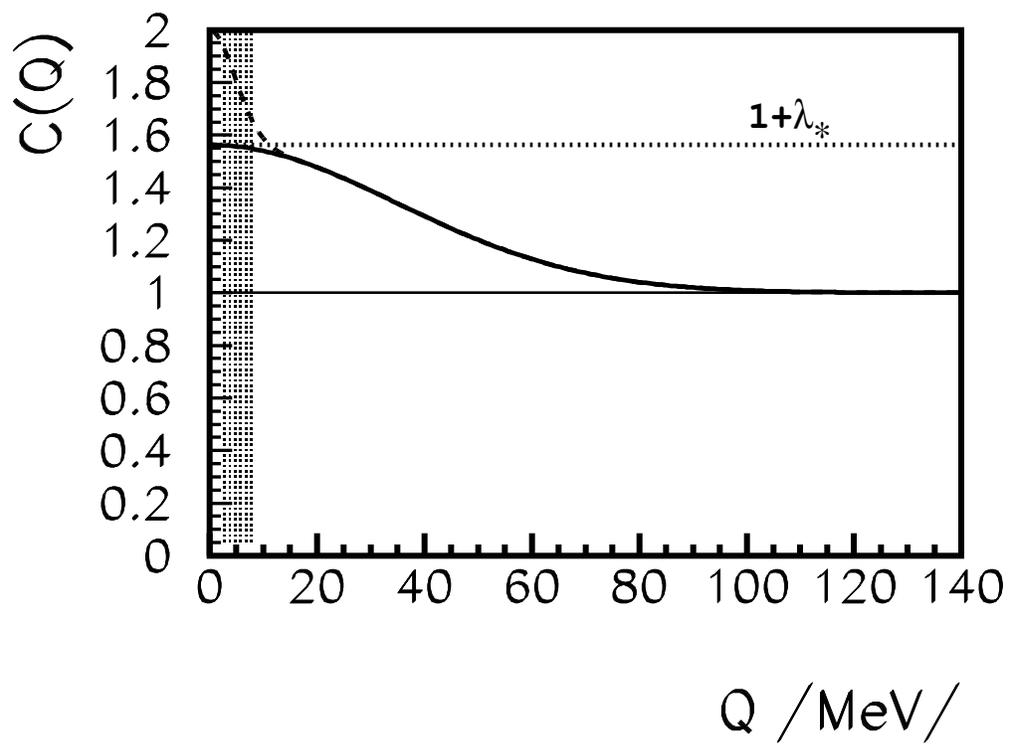
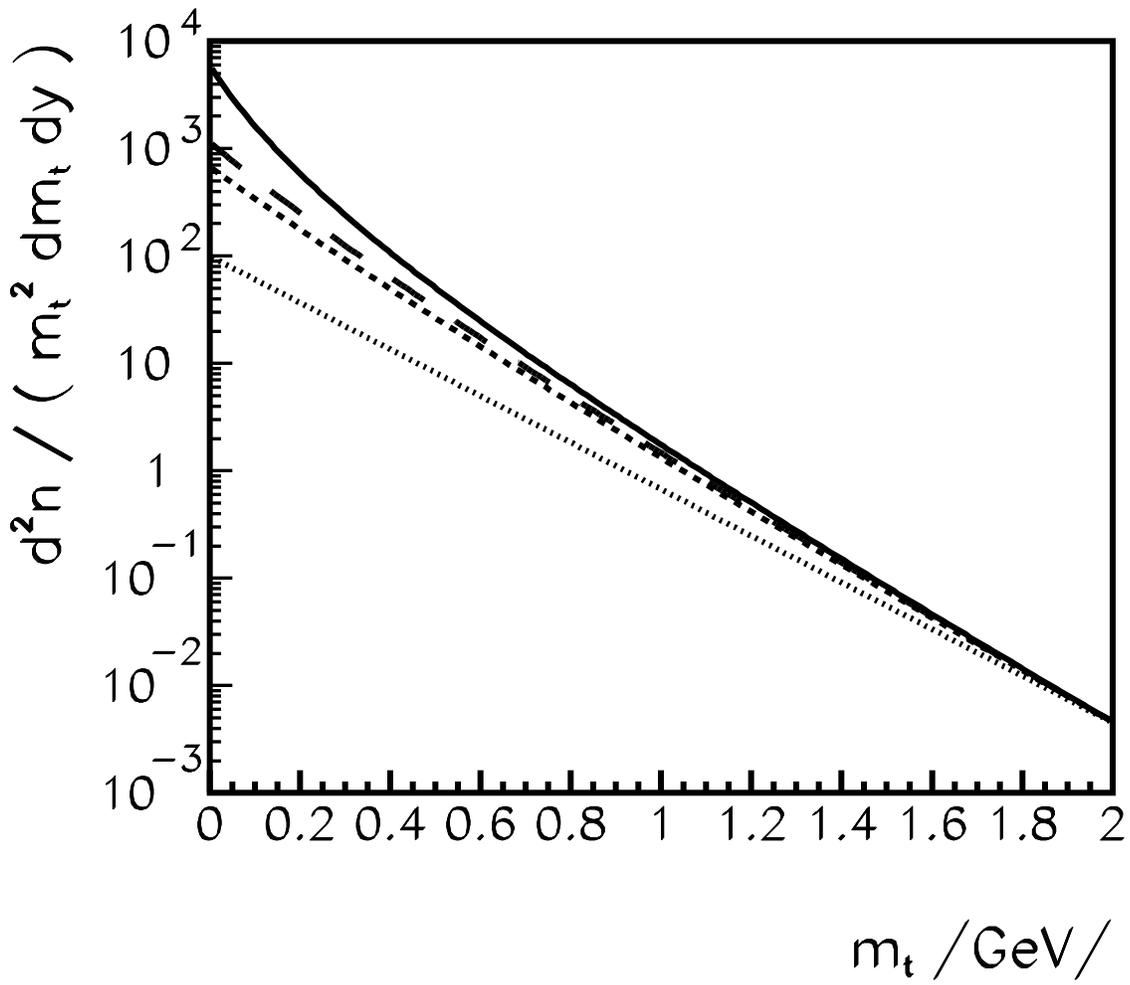
\end{document}